\documentstyle[epsfig,12pt]{article}
\begin{document}

\begin{center}
{\Large \bf Spectral Statistics of RR Intervals in ECG } 
\vspace{0.6cm} \\ 
Mladen MARTINIS, Vesna  MIKUTA-MARTINIS,  Andrea  KNE\v ZEVI\' C,
and Josip \v CRNUGELJ

\vspace{0.6cm}
 {\it  Division of Theoretical Physics \\ Rudjer Bo\v skovi\' c
Institute, Zagreb, Croatia}
\end{center}

\vspace*{0.5cm}

\begin{small}
 The statistical properties (fluctuations) of  heartbeat  intervals (RR intervals) in ECG
 are studied and compared  with the predictions of  Random Matrix Theory (RMT).
It is  found that heartbeat intervals only locally exhibit  the
fluctuation patterns (universality)
 predicted by the RMT.
This finding shows that heartbeat dynamics is of the mixed type
where  regular and irregular (chaotic)  regimes coexist and the
Berry-Robnik theory can be applied. It is also observed that the
distribution of heartbeat intervals is well described by the
one-parameter Brody distribution. The parameter $\beta $ of the
Brody distribution is seen to be connected with the dynamical
state of the heart.

\end{small}

\vspace*{1cm}
\section{\ Introduction}

The time series of heartbeat intervals (RR intervals), used in
various analyses, are usually Holter type data or data from steady
state ambulatory measurements.
The great wealth of data about the dynamics of the heart that is
contained in such ECG records is usually reduced to characterize
only the mean heart rate and the presence and frequency of some
abnormal electrocardiographic complexes. The analysis of
short-range and long-range fluctuations and its universality are
largely ignored.$^{1)}$ The  normal human heartbeats display
considerable variability and nonstationarity over time. The result
is that fluctuations of heartbeat intervals around the mean value
are irregular and complex, even under resting conditions, Fig. 1.
A number of new methods have been proposed to quantify these
complex fluctuations, mainly based on the nonlinear and fractal
analysis. $^{2), 3), 4), 5)}$

In this paper we examine the possibility that the distribution of
heartbeat intervals exhibits universality of the type predicted by
the RMT. $^{6)}$

\section{Statistical analysis of heartbeat intervals}

The heartbeat time series of an human  represents  an ordered
sequence of R-waves (QRS-complexes) occuring at times $\{t_1,
t_2,\cdots ,t_i,\cdots \}$. In this form the heartbeat dynamics
can be regarded as a fractal stochastic point process on a
time-line. The point process is completely described by the set
$\{t_i\}$, or equivalently by the set of  RR-intervals defined as
$\{RR(i) = t_{i+1} - t_i , \,\,\, i=1,2, \cdots \}$.

To see the universality in beat-to-beat RR interval fluctuations
we need to magnify the heartbeat spectrum $\{t_i\}$ so that mean
spacing is unity.
 Let us first determine  the number of  heartbeats up to time $t$.
 It is given by the staircase function  $N(t)$:
\begin{equation}
N(t) = \sum_{i}\theta (t - t_i)
\end{equation}
where $\theta (t)$ is the step function.

In order to be able to compare statistical properties of heartbeat
spectra, in particular the fluctuations around the mean value,
with the predictions of the RMT it is necesary to normalize
(unfolde) the time spectrum $\{t_i\}$ to $\{\tau _i\}$ in such a
way  that the mean spacing between neighbouring $\tau _i$ is
unity. This is achieved by the mapping
\begin{equation}
\tau_i = \bar{N}(t_i), \,\,\, S_i = \tau_{i+1} - \tau_{i}
\end{equation}
where $\bar{N}(t)$ denotes the smoothed staircase which account
for the mean trend in $N(t)$ and $S_i$ is the normalized heartbeat
spacing. The "local" mean spacing is  defined as $\Delta(t) =
[d\bar{N}(t)/dt]^{-1}$ so that $S_i$ becomes
\begin{equation}
S_i = \frac{t_{i+1}-t_{i}}{\Delta[(t_{i+1}+t_{i})/2]}.
\end{equation}
During the real time measurements, we always deal with the finite
number of heartbeats, say $n$. In that case, we  define the
"global" mean spacing as

\begin{equation}
\Delta  = \frac{1}{n-1}\sum_{i=1}^{n-1} RR(i) = \frac{t_n -
t_1}{n-1}
\end{equation}
so that $S_i = RR(i)/\Delta $.

There are two important statistical measures that we shall use to
characterize the heartbeat spectra $\{\tau_i\}$. One is the nearest-neighbor
heartbeat spacing distribution $P(S)dS $ which gives the probability to find
$\tau_i$ in the interval ($S,S + dS$). $P(S)$ is a short range
statistics, it measures the short range correlations between heartbeats.
Its behaviour  for small spacing, $S\rightarrow 0$,
\begin{equation}
P(S) \approx const \times S^{\beta}
\end{equation}
determines the type of short range correlations between heartbeat intervals.
Two cases can be distinguished: a)  when $\beta = 0$ meaning that  the short range
 fluctuations of $\{S_i\}$ are just those of uncorrelated random numbers,
 i.e. Poissonian  with a well-known phenomenom of attraction (clustering),
 and b) when $\beta \neq 0$ meaning that the long range fractal fluctuations
 are present in  the sequence of $S_i$  that is characterized by the phenomenom of
 repulsion. The universal statistics (no free parameters)
 predicted by the RMT $^{6)}$ are given by
 \begin{eqnarray}
 P_{Poisson}(S)   & = &  exp(- S),  \; \;  \beta = 0, \nonumber \\
 P_{GOE}(S)  & = & \frac{\pi }{2} S  exp(-\frac{\pi }{4} S^2), \; \;  \beta = 1, \\
 P_{GUE}(S)  & = & \frac{32}{\pi ^2} S^2 exp(- \frac{4}{\pi } S^2), \; \;  \beta = 2, \nonumber
 \end{eqnarray}
 The GOE  (Gaussian orthogonal ensemble) distribution, known as a
 Wigner distribution, should be valid for irregular (chaotic) systems
 with an antiunitary symmetry, whereas GUE (Gaussian Unitarian Ensemble) should apply
 if the system has no such symmetry.

 To avoid the smoothing procedure on a measured data,
 it is easier and less ambiguous to study the inegrated or the cumulative RR intervals
 \begin{equation}
 I(S) = \int_{0}^{S} P(S) dS
 \end{equation}
 which yields a useful statistics even for small sample of  RR intervals. This gives

 \begin{equation}
 I_{Poisson}(S) = 1 - exp(- S), \; \;  I_{Wigner}(S) = 1 - exp(- \frac{\pi }{4}S^2),
 \end{equation}
 and a nonanalytical expression for $I_{GUE}(S)$.

 Another important  statistics, not considered here,  is the spectral rigidity
 $\Delta _3$  of Dyson and Metha $^{7)}$  which characterizes the long-range
 correlations of heartbeats. It can be related to the Hurst
 exponent $^{8), 9)}$ associated with  the heartbeat time series.

\section{Results and discussion}

Time series of RR intervals were measured at the Institute for
Cardiovascular Disease and Rehabilitation in Zagreb. Each
masurement of ECG was over a time duration of about 15 min.
($\simeq$ 2000 heartbeats), in a controlled ergometric regime.
 These type of measurements are  used as routine in everyday clinical
 diagnostic practice,
because some heart deseases, such as stable angine pectoris (SAP),
usually become transparent under physical activities.

The ECG data were digitized by the WaveBook 512 (Iotech. Cal.
USA), and transferred to a computer. The RR interval series was
passed through a filter that eliminates noise and artefacts. All
R-wave peaks were first edited automatically, after which a
careful manual editing was performed by visual inspection of the
each RR interval. After this, all questionable portions were
excluded manually, and only segments with $>90\%$ sinus beats were
included in the final analysis.

Each measurement consisted of stationary state part (pretrigger
Pt), a few stages of running (P1-P4) and a period of relaxation
(Re), Fig. 1.
\begin{figure}
\epsfig{file=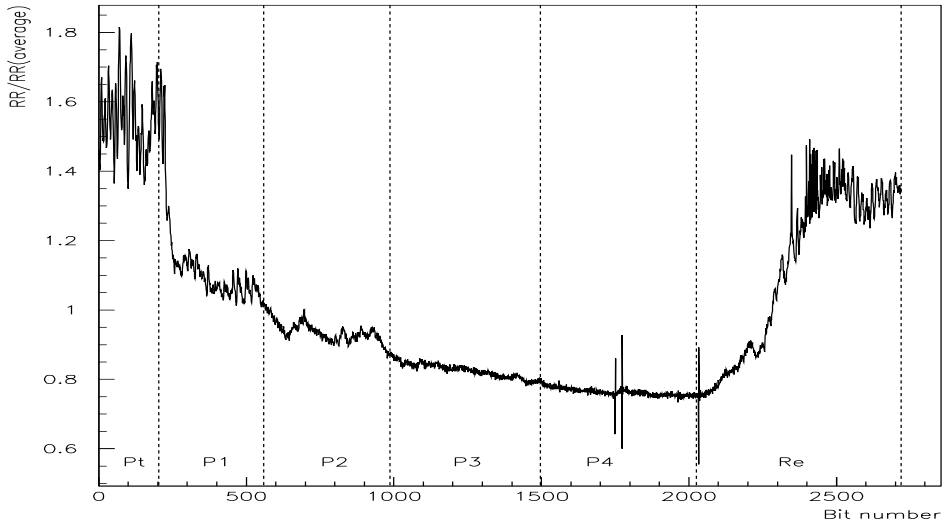,width=14cm,height=8cm}
\caption{Time series of RR intervals as a function of beat number from a
person during an ergometric measurement.}
\end{figure}
 Here we report on the universality analysis of RR interval fluctuations
 in the pretrigger and P1 period only. $^{10)}$ Our patients were divided 
 in two groups:
 one with the evidence of ishemic ST-segment depression (SAP subjects),
 and the control group of healthy subjects. Selection of subjects was
performed by a cardiologist according to the generally accepted
medical knowledge.

The integrated RR interval distributions $I(S)$ are shown in Fig.2
together with $I_W (S)$. 
\begin{figure}
\center
\epsfig{file=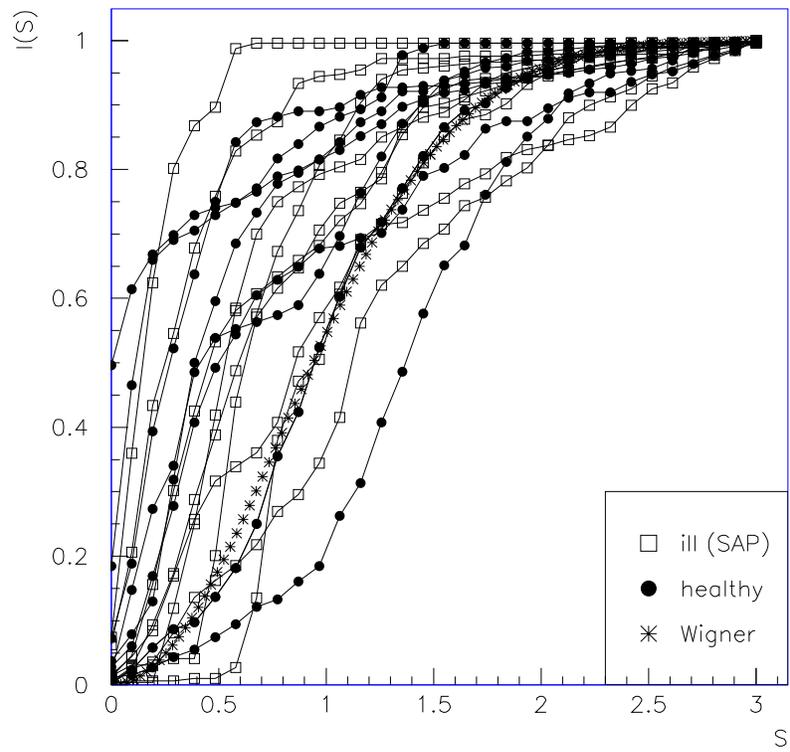,width=15cm}
\caption{Integrated RR interval distributions $I(S)$, in P1 period, in 
comparison with Wigner distribution.}
\end{figure}
Clearly the universal, parameter free,
distributions $P_{Poisson}, P_{W}$ and $P_{GUE}$ of RMT are
excluded. A possible reason for this is that heart is a very
complex  system of nonlinearly coupled biological oscillators
whose motion is partly regular and partly irregular (chaotic) or
mixed . Another possibility is that universality is only
local.$^{10)}$

The first theoretical step towards the understanding of the
regular-irregular coexistence in a dynamical system was the work
by Berry and Robnik $^{11)}$ in which they offered a semiclassical
model for $P(S)$ known as the Berry-Robnik formulae. It is to be
applied at large intervals $S>1$. At small and intermediate
intervals $S<1$ the  Brody distribution $^{12)}$
\begin{equation}
P_{B}(S) = aS^{\beta }exp(- bS^{\beta + 1}), \; \; a= (\beta +1) b,
\; \; b = \{\Gamma (\frac{\beta +2}{\beta +1})\}^{\beta +1}
\end{equation}
gives better fitting results. The Brody distribution becomes
Poisson at $\beta = 0$ and Wigner for $\beta = 1$.

Figure 3 shows that the Brody distribution fits very well the ECG
measured distribution of RR intervals. Each subject has its own
value of $\beta $ which might be an indicator of a dynamical state
of the heart. Indeed,  Fig. 4 clearly shows that $\beta $ could be
a useful parameter for separating SAP subjects from healthy ones.
\begin{figure}
\center
\epsfig{file=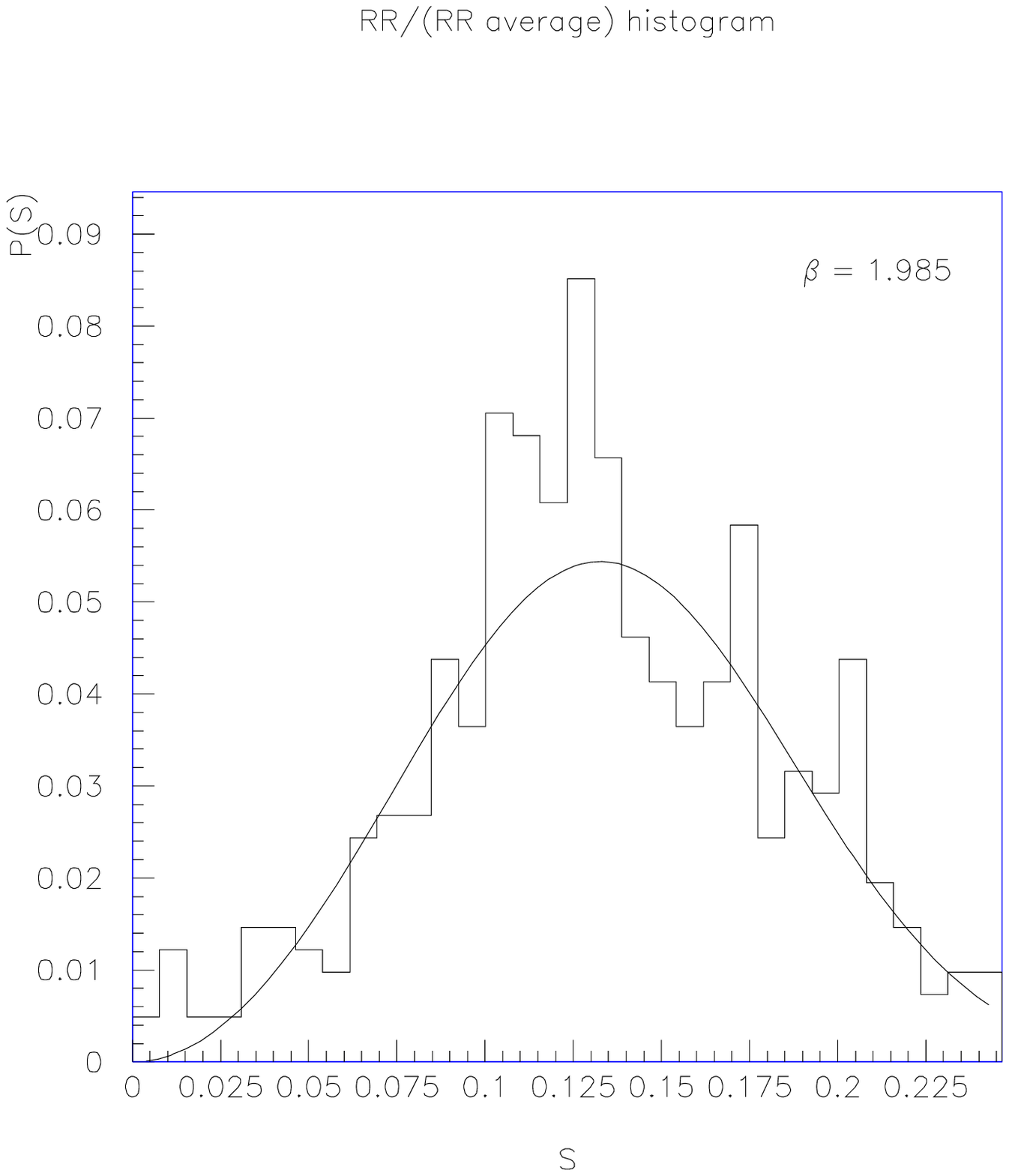,width=6cm} \\
\epsfig{file=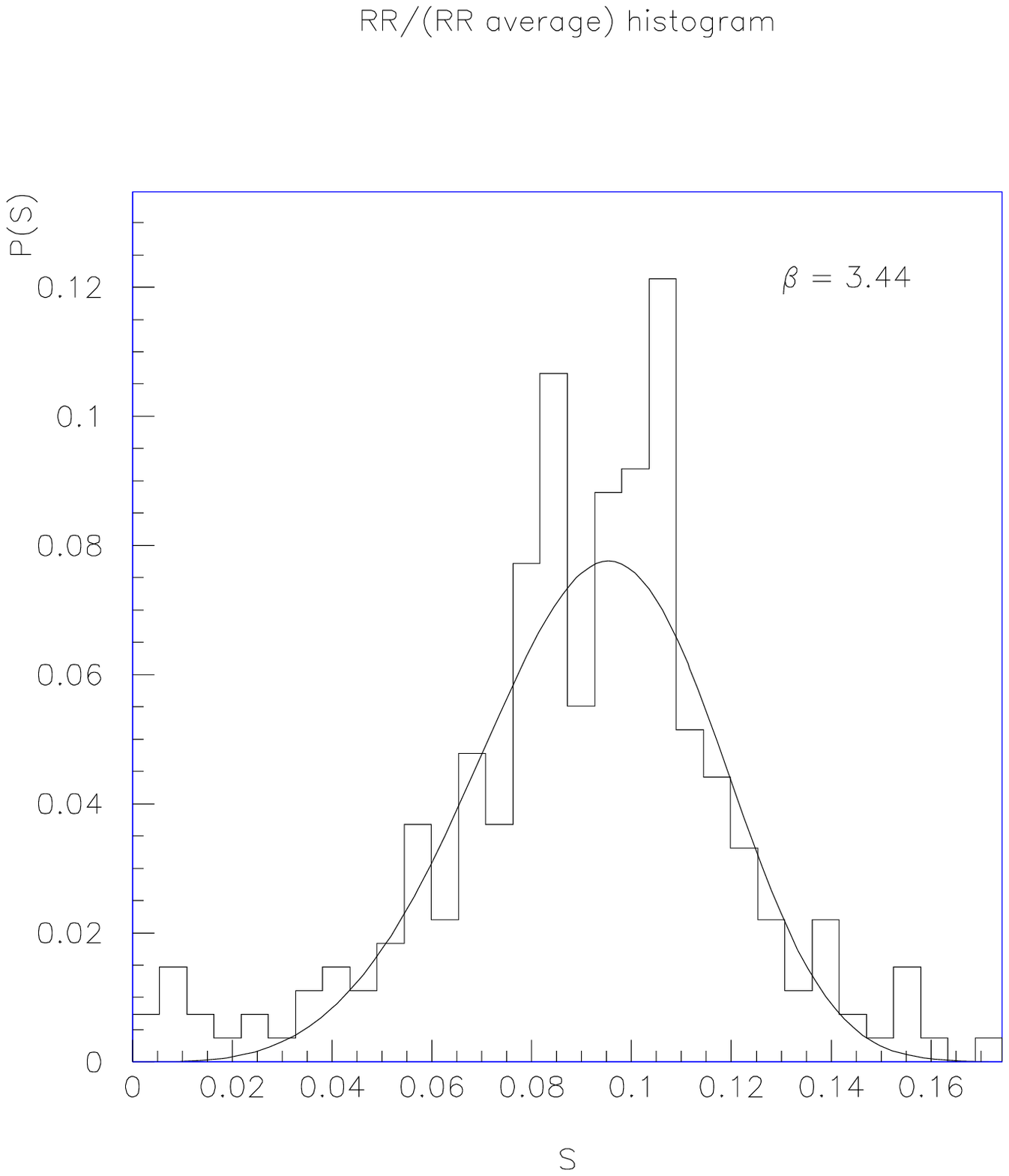,width=6cm} \\
\epsfig{file=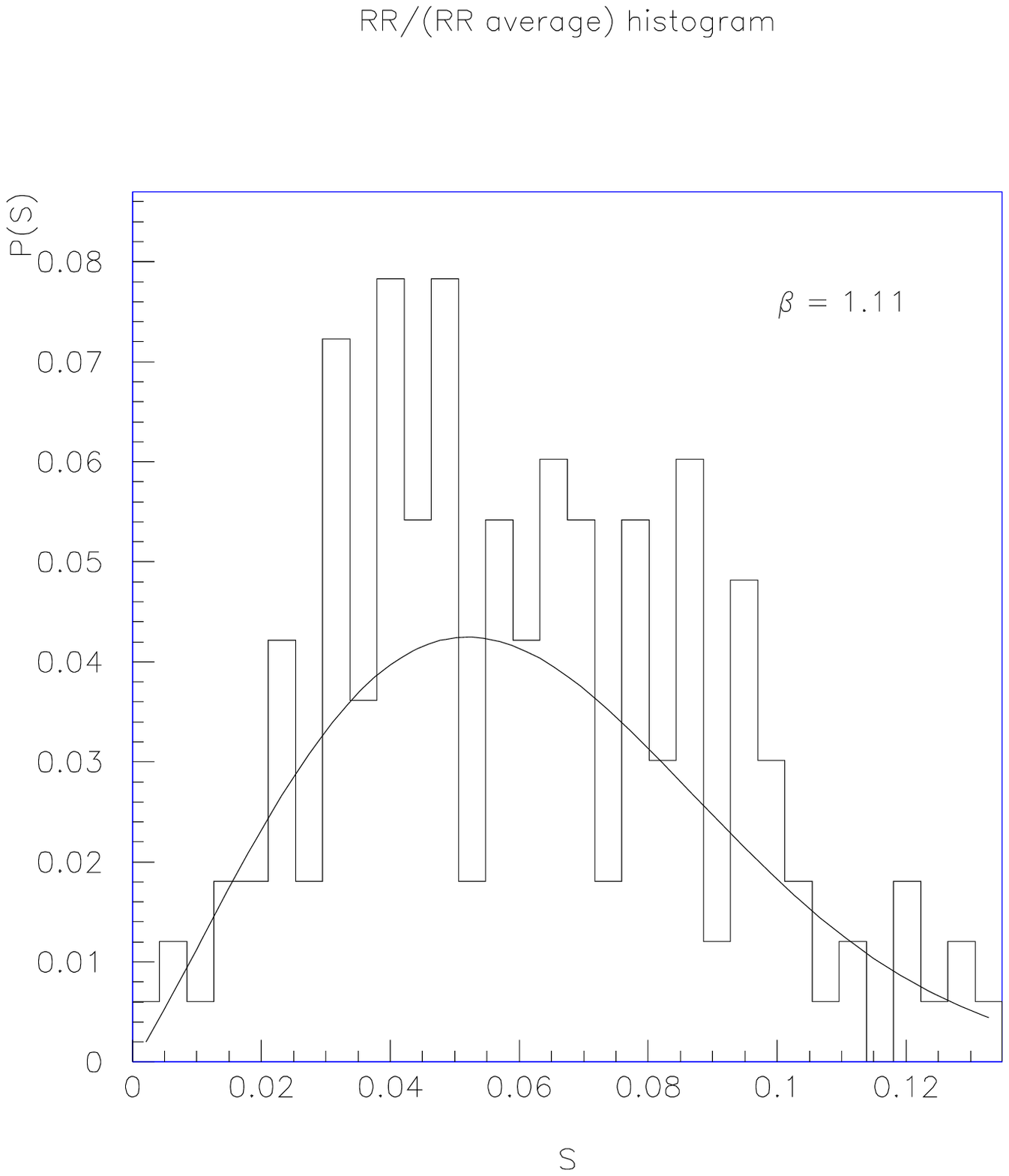,width=6cm}
\caption{RR interval distributions, in Pt period, from three different 
persons in comparison with the one parameter Brody distribution.
[$S=(RR-RR_{min})/\overline{RR}$]}
\end{figure}

\begin{figure}
\center
\epsfig{file=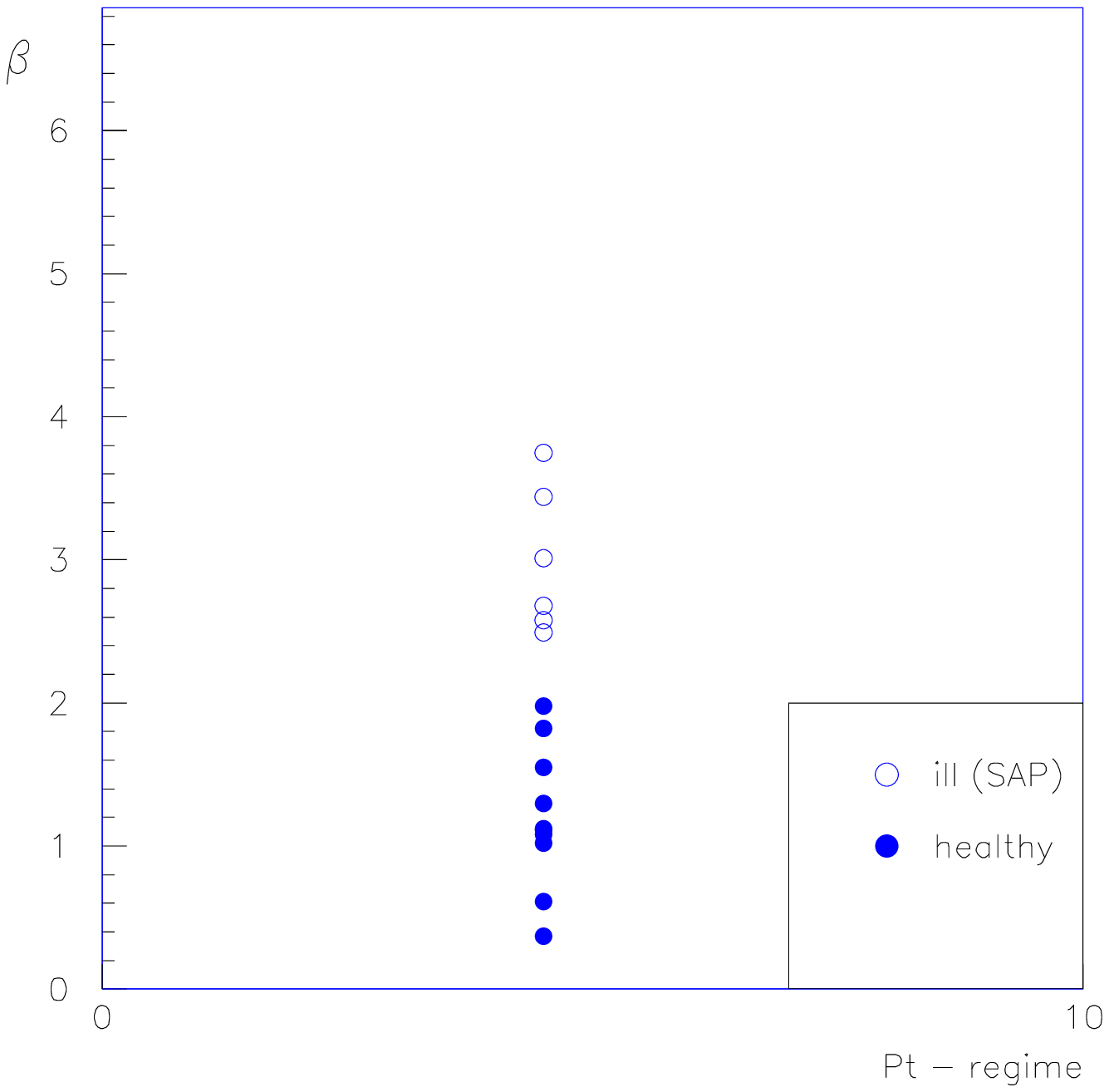,width=15cm}
\caption{Brody parameter $\beta $ as a function  of  a  healthy state of the
heart.}
\end{figure}

\section{Conclusion}

In this paper we have investigated the possibility that
fluctuations of RR intervals exhibit universal statistical laws.
It is found, Fig. 2, that this is not so. Heart is a complex
nonlinear system with many coexisting regular and irregular
motions which manifest themselves through the nonuniversal
fluctuations of RR intervals. It is also found, Fig. 3 and 4, that
the one-parameter Brody distribution could be used successfuly to
describe the fluctuation pattern of RR intervals. The parameter
$\beta $ of the Brody distribution is seen, Fig. 4,to be connected
with the dynamical state of the heart.

Further studies in larger populations are needed to confirm
these interesting results.

\section*{Acknowledgements}
This work was supported by the Ministry of Science and Technology
of the Republic of Croatia.


\begin{thebibliography}{99}
\bibitem{10}
 Programs $P1 - P4$  include controled physical activity  on
the moving belt. Interesting transitory phenomenae between Poisson
and Wigner type distributions are  observed. Detailed analysis
will be reported elsewhere.
\end{thebibliography}
\end{document}